\documentclass[atmp]{ipart_v1}

\Vol{00}
\Issue{0}
\Year{2021}
\firstpage{1}

\usepackage{t1enc}
\usepackage[latin1]{inputenc}
\usepackage[english]{babel}

\usepackage{amsthm}
\usepackage{yfonts}

\usepackage{bbm}
\usepackage{bm}
\usepackage{mathrsfs}

\newcommand{\be}[0]{\begin{equation}}
\newcommand{\ee}[0]{\end{equation}}

\numberwithin{equation}{section}

\theoremstyle{plain}

\begin{document}

\title[The quantum black hole as a gravitational hydrogen atom]{The quantum black hole as a gravitational hydrogen atom}

\author[Christian Corda and Fabiano Feleppa]{Christian Corda and Fabiano Feleppa}

\begin{abstract}
Nathan Rosen's quantization approach to the gravitational collapse
is applied in the simple case of a pressureless \textquotedblleft \emph{star
of dust}\textquotedblright{} by finding the gravitational potential,
the Schr\"odinger equation and the solution for the collapse's energy
levels. By applying the constraints for a Schwarzschild black hole
(BH) and by using the concept of \emph{BH effective state}, previously
introduced by one of the authors (CC), the BH quantum gravitational potential, Schr\"odinger equation and the BH energy spectrum are found. Remarkably, such an energy spectrum
is in agreement (in its absolute value) with the one which was conjectured by Bekenstein
in 1974 and consistent with other ones in the literature.
This approach also allows us to find an interesting
quantum representation of the Schwarzschild BH ground state at the
Planck scale. Moreover, two fundamental issues about black hole quantum physics are addressed by this model: the area quantization and the singularity resolution. As regards the former, a result similar to the one obtained by Bekenstein, but with a different coefficient, has been found. About the latter, it is shown that the traditional classical singularity in the core of the Schwarzschild BH is replaced, in a full quantum treatment, by a two-particle system where the two components strongly interact with each other via a quantum gravitational potential. The two-particle system seems to be non-singular from the quantum point of view and is analogous to the hydrogen atom because it consists of a \textquotedblleft nucleus\textquotedblright{} and an \textquotedblleft electron\textquotedblright{}.
\end{abstract}

\maketitle

\section{Introduction}

It is a general conviction that, in the search of a quantum gravity
theory, a black hole should play a role similar to that of the hydrogen atom in quantum
mechanics \cite{key-9}. It should be a ``theoretical laboratory''
where one discusses and tries to understand conceptual problems and
potential contradictions in the attempt to unify Einstein's general
theory of relativity with quantum mechanics. This analogy suggests
that black holes should be regular quantum systems with a discrete mass spectrum
\cite{key-9}. In this paper, the authors attempt to contribute
to the above by finding the Schr\"odinger equation and the wave
function of the Schwarzschild BH. The knowledge of such quantities
could, in principle, also play a role in the solution of the famous
BH information paradox \cite{key-10} because here black holes seem to be well defined quantum mechanical systems, having
ordered and discrete quantum spectra. This issue appears consistent with
the unitarity of the underlying quantum gravity theory and with the
idea that information should come out in BH evaporation. 

A quantization approach proposed 25 years ago by the historical collaborator of Einstein,
Nathan Rosen \cite{key-5}, will be applied to the gravitational collapse in the simple case of a pressureless \textquotedblleft \emph{star
of dust}\textquotedblright. Therefore, the gravitational potential, the
Schr\"odinger equation and the solution for the collapse's energy levels
will be found. After that, the constraints for a BH will be applied
and this will permit to find the BH quantum gravitational potential, Schr\"odinger equation and energy spectrum. Such an energy spectrum, in its absolute value, is in agreement with both the one conjectured by J. Bekenstein in 1974 \cite{key-7} and that found by Maggiore's description of black hole in terms of quantum membranes \cite{key-8}. Rosen's approach also allows us to find an interesting quantum representation of the Schwarzschild BH ground state at the Planck scale.

It is well-known that the canonical quantization of general relativity leads to the Wheeler-DeWitt equation introducing the so-called Superspace, an infinite-dimensional space of all possible 3-metrics; Rosen, instead, preferred to start his work from the classical cosmological equations using a simplified quantization scheme, reducing, at least formally, the cosmological Einstein-Friedman equations of general relativity to a quantum mechanical system; the Friedman equations can be then recast as a Schr\"odinger equation and the cosmological solutions can be read as eigensolutions of such a \textquotedblleft cosmological Schr\"odinger equation. In this way Rosen found that, in the case of a Universe filled with pressureless matter, the equation is like that for the $s$ states of a hydrogen-like atom \cite{key-5}. It is important to recall that quantization of FLRW universe date back at least to DeWitt's famous 1967 paper \cite{key-42}, where one understands that a large number of particles is required in order to ensure the semiclassical behaviour.

Furthermore, we try to clarify two important issues such as Bekenstein area law and singularity resolution. With regard to the former, a result similar to that obtained by Bekenstein, but with a different coefficient, has been found. About the latter, it is shown that the traditional classical singularity in the core of the Schwarzschild BH is replaced,
in a full quantum treatment, by a two-particle system where the two components strongly interact with each other via a quantum gravitational potential. The two-particle system seems to be non-singular from the quantum point of view and is analogous to the hydrogen atom because it consists of a \textquotedblleft core\textquotedblright{} and an \textquotedblleft electron\textquotedblright{}. 

Returning to the Rosen's quantization approach, it has also been recently applied to a cosmological framework by one of the authors
(FF) and collaborators in \cite{key-11} and to the famous Hartle-Hawking
initial state by both of the authors and I. Licata in \cite{key-26}.

For the sake of completeness, one stresses that this quantization
approach is only applicable to homogeneous space-times where the Weyl
tensor vanishes. As soon as one introduces inhomogeneities (for example in the
Lema\^itre-Tolman-Bondi (LTB) models \cite{key-23,key-24,key-25}), there exists open
sets of initial data where the collapse ends to a BH absolutely similar
to Oppenheimer-Snyder-Datt (OSD) collapse. Hence, the exterior space-time
remains the same whereas the interior is absolutely different. A further
generalization of the proposed approach, which could
be the object of future works, should include also a cosmological
term or other sources of dark energy.
One also underlines that the Oppenheimer and Snyder gravitational collapse is not physical and can be considered a toy model. But here the key point is that, as it is well known from the historical paper of Oppenheimer and Snyder \cite{key-2}, the final state of this simplified gravitational collapse is the SBH, which, instead, has a fundamental role in quantum gravity. It is indeed a general conviction, arising from an idea of Bekenstein \cite{key-9}, that, in the search of a quantum gravity theory, the SBH should play a role similar to the hydrogen atom in quantum mechanics. Thus, despite non-physical, the Oppenheimer and Snyder gravitational collapse must be here considered as a tool which allows to understand a fundamental physical system, that is the SBH. In fact, it will be shown that, by setting the constrains for the formation of the SBH in the quantized Oppenheimer and Snyder gravitational collapse, one arrives to quantize the SBH, and this will be a remarkable, important result in the quantum gravity's search.

\section{Application of Rosen's quantization approach to the gravitational
collapse}

Classically, the gravitational collapse in the simple case of a pressureless
\textquotedblleft \emph{star
of dust}\textquotedblright{} with uniform density
is well known \cite{key-1}. Historically, it
was originally analysed in the famous paper of Oppenheimer and Snyder
\cite{key-2}, while a different approach has been developed by
Beckerdoff and Misner \cite{key-3}. Furthermore, a non-linear electrodynamics
Lagrangian has been recently added in this collapse's framework by one of the
authors (CC) and Herman J. Mosquera Cuesta in \cite{key-4}. This different approach
allows to find a way to remove the black hole singularity at the classical
level. The traditional, classical framework of this
kind of gravitational collapse is well known \cite{key-1,key-2,key-3}. In the following we will follow \cite{key-1}.  In regard to the
interior of the collapsing star, it is described by the well-known Friedmann-Lema\^itre-Robertson-Walker
(FLRW) line-element with comoving hyper-spherical coordinates $\chi,$
$\theta,$ $\varphi$. Therefore, one writes down
(hereafter Planck units will be used, i.e., $G=c=k_{B}=\hbar=\frac{1}{4\pi\epsilon_{0}}=1$)
\begin{equation}
ds^{2}=d\tau^{2}+a^2(\tau)[(-d\chi^{2}-\sin^{2}\chi(d\theta^{2}+\sin^{2}\theta d\varphi^{2})],\label{eq: metrica conformemente piatta}
\end{equation}
where the origin of coordinates is set at the centre of the star,
and the following relations hold:
\begin{equation}
\begin{array}{c}
a=\frac{1}{2}a_{m}\left(1+\cos\eta\right),\\
\\
\tau=\frac{1}{2}a_{m}\left(\eta+\sin\eta\right).
\end{array}\label{eq: cycloidal relation}
\end{equation}
The density is given by
\begin{equation}
\rho=\left(\frac{3a_{m}}{8\pi}\right)a^{-3}=\left(\frac{3}{8\pi a_{m}^{2}}\right)\left[\frac{1}{2}\left(1+\cos\eta\right)\right]^{-3}.\label{eq: density}
\end{equation}
Setting $\sin^{2}\chi$ one chooses the case of positive curvature,
which corresponds to a gas sphere whose dynamics begins at rest with
a finite radius and, in turn, it is the only one of interest.
Thus, the choice $k=1$ is made for dynamical reasons (the initial
rate of change of density is null, that means ``momentum of maximum
expansion''), but the dynamics also depends on the
field equations. As it has been stressed in the Introduction,
for isotropic models, a cosmological term - or other sources of dark
energy - can be in principle included in future works, in order to
obtain a more realistic physical framework for the collapse. 

In order to discuss the simplest model of a \emph{``star of dust''},
that is the case of zero pressure, one sets the stress-energy tensor
as
\begin{equation}
T=\rho u\otimes u,\label{eq: stress energy}
\end{equation}
where $\rho$ is the density of the collapsing star and $u$ the four-vector
velocity of the matter. On the other hand, the external geometry is
given by the Schwartzschild line-element
\begin{equation}
ds^{2}=\left(1-\frac{2M}{r}\right)dt^{2}-r^{2}\left(\sin^{2}\theta d\varphi^{2}+d\theta^{2}\right)-\frac{dr^{2}}{1-\frac{2M}{r}},\label{eq: Hilbert}
\end{equation}
where $M$ is the total mass of the collapsing star. Since there are
no pressure gradients, which can deflect the particles motion, the
particles on the surface of any ball of dust move along radial geodesics
in the exterior Schwarzschild space-time. Considering
a ball which begins at rest with finite radius (in terms of the Schwarzschild
radial coordinate) $r=r_{i}$ at the (Schwarzschild) time $t=0,$
the geodesics motion of its surface is given by the following equations: 
\begin{equation}
r=\frac{1}{2}r_{i}\left(1+\cos\eta\right),\label{eq: geodesics surface radius}
\end{equation}
\vspace{-0.8cm}
\begin{equation}
\begin{array}{c}
\hspace{0.55cm} t=2M\ln\left[\frac{\sqrt{\frac{r_{i}}{2M}-1}+\tan\left(\frac{\eta}{2}\right)}{\sqrt{\frac{r_{i}}{2M}-1}-\tan\left(\frac{\eta}{2}\right)}\right]\\
\\
\quad\quad\quad\quad\quad\;\:\,\hspace{1.5cm}+2M\sqrt{\frac{r_{i}}{2M}-1}\left[\eta+\left(\frac{r_{i}}{4M}\right)\left(\eta+\sin\eta\right)\right].
\end{array}\label{eq: geodesic surface time}
\end{equation}
The proper time measured by a clock put on the surface of the collapsing
star can be written as
\begin{equation}
\tau=\sqrt{\frac{r_{i}^{3}}{8M}}\left(\eta+\sin\eta\right).\label{eq: tau}
\end{equation}
The collapse begins at $r=r_{i},$ $\eta=\tau=t=0,$ and ends
at the singularity $r=0,$ $\eta=\pi$ after a duration of proper
time measured by the falling particles
\begin{equation}
\Delta\tau=\pi\sqrt{\frac{r_{i}^{3}}{8M}},\label{eq: delta tau}
\end{equation}
which coincidentally corresponds, as it is well known, to the interval
of Newtonian time for free-fall collapse in Newtonian theory. Different
from the cosmological case, where the solution is homogeneous and
isotropic everywhere, here the internal homogeneity and isotropy of
the FLRW line-element are broken at the star's surface, that is, at
some radius $\chi=\chi_{0}$. At that surface, which
is a 3-dimensional world tube enclosing the star's fluid,
the interior FLRW geometry must smoothly match the exterior Schwarzschild
geometry. One considers a range of $\chi$ given by
$0\leq\chi\leq\chi_{0}$, with $\chi_{0}<\frac{\pi}{2}$ during the
collapse. For the pressureless case the match is possible. The external Schwarzschild solution indeed predicts
a cycloidal relation for the star's circumference
\begin{equation}
\begin{array}{c}
C=2\pi r=2\pi\left[\frac{1}{2}r_{i}\left(1+\cos\eta\right)\right],\\
\\
\tau=\sqrt{\frac{r_{i}^{3}}{8M}}\left(\eta+\sin\eta\right).
\end{array}\label{eq: Circonferenza esterna}
\end{equation}
The interior FLRW predicts a similar cycloidal relation
\begin{equation}
\begin{array}{c}
C=2\pi r=2\pi a\sin\chi_{0}=\pi\sin\chi_{0}a_{m}\left(1+\cos\eta\right),\\
\\
\tau=\frac{1}{2}a_{m}\left(\eta+\sin\eta\right).
\end{array}\label{eq: circonferenza esterna}
\end{equation}
Therefore, the two predictions agree perfectly for all time if and
only if
\begin{equation}
\begin{array}{c}
r_{i}=a_{0}\sin\chi_{0},\\
\\
\hspace{0.25cm}M=\frac{1}{2}a_{0}\sin^{3}\chi_{0,}
\end{array}\label{eq: matching}
\end{equation}
where $r_{i}$ and $a_{0}$ represent the values of the Schwarzschild radial
coordinate in Eq. (\ref{eq: Hilbert}) and of the scale factor in
Eq. (\ref{eq: metrica conformemente piatta}) at the beginning of
the collapse, respectively. Thus, Eqs. (\ref{eq: matching}) represent
the requested match, while the Schwarzschild radial coordinate, in
the case of the matching between the internal and external geometries,
is
\begin{equation}
r=a\sin\chi_{0}.\label{eq: mach}
\end{equation}
The attentive reader notes that the initial conditions on the matching
are the simplest possible that could be relaxed, still having a continuous
matching without extra surface terms. In fact, taking the interior
solution to be homogeneous requires very fine tuned initial conditions
for the collapse and the dynamics of the edge. So, on one hand,
further analyses for a better characterization of the initial conditions
on the matching between the internal and external geometries could
be the object of future works. On the other hand, despite the analysis
of this paper is not as general as possible, one stresses that the BH quantization is one of the most important problems of modern theoretical physics which has not yet been solved.
Thus, in order to attempt to solve such a fundamental problem, one
must start from the simplest case rather than from more complicated
ones. This is in complete analogy with the history of general relativity. In fact, the first solution of Einstein field equations was the Schwartzschild
solution, but it was not a general, rotating solution which included
cosmological term or other sources of dark energy, as well the corresponding
gravitational collapse developed by Oppenheimer and Snyder did not
include a class of non-homogeneous models. Thus, as this is a new
approach to the BH quantization, here one starts from the simplest
conditions rather than from more complicated ones. So, the initial conditions
on the matching that are applied here are exactly the ones proposed
by Oppenheimer and Snyder in their paper on the gravitational
collapse \cite{key-2}. It is well known that the final result of
the gravitational collapse studied by Oppenheimer and Snyder is the
Schwarzschild BH \cite{key-1}. 

In the following, the quantization approach derived by Rosen in \cite{key-5}
will be applied to the above case; some differences will be found,
because here one analyses the case of a collapsing star, while Rosen
analysed a closed homogeneous and isotropic universe.
Let us start by rewriting the FLRW line-element (\ref{eq: metrica conformemente piatta})
in spherical coordinates and comoving time as \cite{key-1,key-5} 
\begin{equation}
ds^{2}=d\tau^{2}-a^{2}(\tau)\left(\frac{dr^{2}}{1-r^{2}}+r^{2}d\theta^{2}+r^{2}\sin^{2}\theta d\varphi^{2}\right).\label{eq: FLRW}
\end{equation}
The Einstein field equations
\begin{equation}
G_{\mu\nu}=-8\pi T_{\mu\nu}\label{eq: Einstein field equation}
\end{equation}
gives the relations (we are assuming zero pressure)
\begin{equation}
\begin{array}{c}
\dot{a}^{2}=\frac{8}{3}\pi a^{2}\rho-1,\\
\\
\ddot{a}=-\frac{4}{3}\pi a\rho,
\end{array}\label{eq: evoluzione}
\end{equation}
with $\dot{a}=\frac{da}{d\tau}$. For consistency, one gets
\begin{equation}
\frac{d\rho}{da}=-\frac{3\rho}{a},\label{eq: consistenza}
\end{equation}
which, when integrated, gives
\begin{equation}
\rho=\frac{C}{a^{3}}.\label{eq: densit=0000E0}
\end{equation}
In the case of a collapse $C$ is determined by the initial conditions which predict a cycloidal relation for the star's circumference, see Eqs. (\ref{eq: circonferenza esterna}) and Section 32.4 in Ref. \cite{key-1}. One gets
\begin{equation}
C=\frac{3a_{0}}{8\pi}.\label{eq: C}
\end{equation}
This value is consistent with the one found by Rosen \cite{key-5}. 
Thus, one rewrites Eq. (\ref{eq: densit=0000E0}) as 
\begin{equation}
\rho=\frac{3a_{0}}{8\pi a^{3}}.\label{eq: densit=0000E0 2}
\end{equation} 
Multiplying the first of (\ref{eq: evoluzione}) by $M/2$ one
gets
\begin{equation}
\frac{1}{2}M\dot{a}^{2}-\frac{4}{3}\pi Ma^{2}\rho=-\frac{M}{2},\label{eq: energy equation for a particle}
\end{equation}
which seems like the energy equation for a particle in one-dimensional
motion having coordinate $a$:
\begin{equation}
E=T+V,\label{eq: energia totale}
\end{equation}
where the kinetic energy is
\begin{equation}
T=\frac{1}{2}M\dot{a}^{2},\label{eq: energia cinetica}
\end{equation}
and the potential energy is
\begin{equation}
V(a)=-\frac{4}{3}\pi Ma^{2}\rho.\label{eq: energia potenziale}
\end{equation}
Thus, the total energy is
\begin{equation}
E=-\frac{M}{2}.\label{eq: energia totale 2}
\end{equation}
From the second of Eqs. (\ref{eq: evoluzione}), one gets the equation
of motion of this particle:
\begin{equation}
M\ddot{a}=-\frac{4}{3}M\pi a\rho.\label{eq: equation of motion}
\end{equation}
The momentum of the particle is
\begin{equation}
P=M\dot{a},\label{eq: momentum}
\end{equation}
with an associated Hamiltonian
\begin{equation}
\mathcal{H}=\frac{P^{2}}{2M}+V.\label{eq: Hamiltonian}
\end{equation}
Till now, the problem has been discussed from the classical point
of view. In order to discuss it from the quantum point of view, one
needs to define a wave-function as
\begin{equation}
\Psi\equiv\Psi\left(a,\tau\right).\label{eq: wave-function}
\end{equation}
Thus, in correspondence of the classical equation (\ref{eq: Hamiltonian}),
one gets the traditional Schr\"odinger equation
\begin{equation}
i\frac{\partial\Psi}{\partial\tau}=-\frac{1}{2M}\frac{\partial^{2}\Psi}{\partial a^{2}}+V\Psi.\label{eq: Schrodinger equation}
\end{equation}
For a stationary state with energy $E$ one obtains
\begin{equation}
\Psi=\Psi\left(a\right)\exp\left(-iE\tau\right),\label{eq: separazione}
\end{equation}
and Eq. (\ref{eq: wave-function}) becomes
\begin{equation}
-\frac{1}{2M}\frac{\partial^{2}\Psi}{\partial a^{2}}+V\Psi=E\Psi.\label{eq: Schrodinger equation 2}
\end{equation}
Inserting Eq. (\ref{eq: densit=0000E0 2}) into Eq. (\ref{eq: energia potenziale})
one obtains 
\begin{equation}
V(a)=-\frac{Ma_{0}}{2a}.\label{eq: energia potenziale 2}
\end{equation}
Setting
\begin{equation}
\Psi=aX,\label{eq: X}
\end{equation}
Eq. (\ref{eq: Schrodinger equation 2}) becomes
\begin{equation}
-\frac{1}{2M}\left(\frac{\partial^{2}X}{\partial a^{2}}+\frac{2}{a}\frac{\partial X}{\partial a}\right)+VX=EX.\label{eq: Schrodinger equation 3}
\end{equation}
With $V$ given by Eq. (\ref{eq: energia potenziale 2}), (\ref{eq: Schrodinger equation 3})
is analogous to the Schr\"odinger equation in polar coordinates for
the $s$ states ($l=0$) of a hydrogen-like atom \cite{key-6}
in which the squared electron charge $e^{2}$ is replaced by $\frac{Ma_{0}}{2}$.
Thus, for the bound states ($E<0$) the energy spectrum is 
\begin{equation}
E_{n}=-\frac{a_{0}^{2}M^{3}}{8n^{2}},\label{eq: spettro energia}
\end{equation}
where $n$ is the principal quantum number. At this point, one inserts (\ref{eq: energia totale 2}) into (\ref{eq: spettro energia}),
obtaining the mass spectrum as 
\begin{equation}
M_{n}=\frac{a_{0}^{2}M_{n}^{3}}{4n^{2}}\Rightarrow M_{n}=\frac{2n}{a_{0}}.\label{eq: spettro massa}
\end{equation}
On the other hand, by using Eq. (\ref{eq: energia totale 2}) one
finds the energy levels of the collapsing star as 
\begin{equation}
E_{n}=-\frac{n}{a_{0}}.\label{eq: energy levels}
\end{equation}
In fact, Eq. (\ref{eq: spettro massa}) represents the spectrum of
the total mass of the collapsing star, while Eq. (\ref{eq: energy levels})
represents the energy spectrum of the collapsing
star where the gravitational energy, which is given by Eq. (\ref{eq: energia potenziale 2}),
is included. The total energy of a quantum system with bound states
is indeed negative. 

What is the meaning of Eq. (\ref{eq: energy levels}) and of its ground state? One sees that the Hamiltonian (\ref{eq: Hamiltonian}) governs the quantum mechanics of the gravitational collapse. Therefore, the square of the wave function (\ref{eq: wave-function}) must be interpreted as the probability density of a single particle in a finite volume. Thus, the integral over the entire volume must be normalized to unity as 
\begin{equation}
\int dx^{3}\left|X\right|^{2}=1.\label{eq: Normalization}
\end{equation}
For stable quantum systems, this normalization must remain the same at all times of the collapse's evolution. As the wave function (\ref{eq: wave-function})
obeys the  Schrodinger equation (\ref{eq: Schrodinger equation}),
this is assured if and only if the Hamiltonian operator (\ref{eq: Hamiltonian})
is Hermitian \cite{key-43}. In other words, the Hamiltonian operator
(\ref{eq: Hamiltonian}) must satisfy for arbitrary wave functions
$X_{1}$ and $X_{2}$ the equality \cite{key-43}
\begin{equation}
\int dx^{3}\left[HX_{2}\right]^{*}X_{1}=\int dx^{3}X_{2}^{*}HX_{1}.\label{eq: hermiticit=0000E0}
\end{equation}
One notes that both $\vec{p}$ and $a$ are Hermitian operators. Therrefore,
the Hamiltonian (\ref{eq: Hamiltonian}) is automatically  a Hermitian
operator because it is a sum of a kinetic and a potential energy \cite{key-43},
\begin{equation}
H=T+V.\label{eq: energia totale}
\end{equation}
This is always the case for non-relativistic particles in Cartesian-like coordinates and works also for the gravitational collapse under consideration. In this framework, the ground state of Eq. (\ref{eq: energy levels}) represents the minimum energy which can collapse in an astrophysical scenario. We conclude that the gravitational collapse can be interpreted as a  ``perfect" quantum system.

It is also important to clarify the issue concerning the
gravitational energy. It is well known that, in the framework of the general theory of
relativity, the gravitational energy cannot be localized \cite{key-1}.
This is a consequence of Einstein's equivalence principle (EEP) \cite{key-1},
which implies that one can always find in any given locality a reference's
frame (the local Lorentz reference's frame) in which all local gravitational
fields are null. No local gravitational fields means no local gravitational
energy-momentum and, in turn, no stress-energy tensor for the gravitational
field. In any case, this general situation admits an
important exception \cite{key-1}, given by the case of a spherical
star \cite{key-1}, which is exactly the case analysed in this paper.
In fact, in this case the gravitational energy is localized not by
mathematical conventions, but by the circumstance that transfer of
energy is detectable by local measures, see Box 23.1 of \cite{key-1}
for details. Therefore, one can surely consider Eq. (\ref{eq: energia potenziale 2})
as the gravitational potential energy of the collapsing star.

\section{Black hole energy spectrum, ground state and singularity resolution}

Thus, let us see what happens when the star is completely collapsed,
i.e. when the star is a BH. One sees that, inserting $r_{i}=2M=r_{g},$
where $r_{g}$ is the gravitational radius (the Schwarzschild radius),
in Eqs. (\ref{eq: matching}), one obtains $\sin^{2}\chi_{0}=1$. Therefore,
as the range $\chi>\frac{\pi}{2}$ must be discarded \cite{key-1},
one concludes that it is $\chi_{0}=\frac{\pi}{2}$, $r=a$ and $r_{i}=a_{0}=2M=r_{g}$
in Eqs. (\ref{eq: matching}) and (\ref{eq: mach}) for a BH.
Then, Eqs. from (\ref{eq: energia potenziale 2}) to (\ref{eq: energy levels})
become, respectively,
\begin{equation}
V(r)=-\frac{M^{2}}{r},\label{eq: energia potenziale BH}
\end{equation}
\begin{equation}
\Psi=rX,\label{eq: X BH}
\end{equation}
\begin{equation}
-\frac{1}{2M}\left(\frac{\partial^{2}X}{\partial r^{2}}+\frac{2}{r}\frac{\partial X}{\partial r}\right)+VX=EX,\label{eq: Schrodinger equation BH}
\end{equation}
\begin{equation}
E_{n}=-\frac{r_{g}^{2}M^{3}}{8n^{2}},\label{eq: spettro energia BH}
\end{equation}
\begin{equation}
M_{n}=\sqrt{n},\label{eq: spettro massa buco nero}
\end{equation}
\begin{equation}
E_{n}=-\sqrt{\frac{n}{4}}.\label{eq: spettro energia buco nero}
\end{equation}
Eqs. (\ref{eq: energia potenziale BH}), (\ref{eq: Schrodinger equation BH}),
(\ref{eq: spettro massa buco nero}) and (\ref{eq: spettro energia buco nero})
should be the exact gravitation potential energy\emph{,} Schr\"odinger
equation, mass spectrum and energy spectrum for the Schwarzschild
BH interpreted as ``gravitational hydrogen atom'', respectively.
Actually, a further final correction is needed. To clarify this point,
let us compare Eq. (\ref{eq: energia potenziale BH}) with the analogous
potential energy of an hydrogen atom, which is \cite{key-6} 
\begin{equation}
V(r)=-\frac{e^{2}}{r}.\label{eq: energia potenziale atomo idrogeno}
\end{equation}
Eqs. (\ref{eq: energia potenziale BH}) and (\ref{eq: energia potenziale atomo idrogeno})
are formally identical, but there is an important physical difference. In the
case of Eq. (\ref{eq: energia potenziale atomo idrogeno}) the electron's
charge is constant for all the energy levels of the hydrogen atom.
Instead, in the case of Eq. (\ref{eq: energia potenziale BH}), based
on the emissions of Hawking quanta or on the absorptions of external
particles, the BH mass changes during the jumps from one energy level
to another. In fact, such a BH mass decreases for emissions and increases
for absorptions. Therefore, one must also consider this dynamical behavior. One way to take into account this dynamical behavior
is by introducing the \emph{BH effective state} (see \cite{key-13,key-14} for details).
Let us start from the emissions of Hawking quanta. If one neglects
the above mentioned BH dynamical behavior, the probability of emission of Hawking
quanta is the one originally found by Hawking, which represents a
strictly thermal spectrum \cite{key-16}
\begin{equation}
\Gamma\sim\exp\left(-\frac{\omega}{T_{H}}\right),\label{eq: hawking probability}
\end{equation}
where $\omega$ is the energy-frequency of the emitted particle
and $T_{H}\equiv\frac{1}{8\pi M}$ is the Hawking temperature. Taking
into account the BH dynamical behavior, i.e., the BH contraction allowing
a varying BH geometry, one gets the famous correction found by Parikh and
Wilczek \cite{key-17}:
\begin{equation}
\Gamma\sim\exp\left[-\frac{\omega}{T_{H}}\left(1-\frac{\omega}{2M}\right)\right]\quad\Longrightarrow\quad\Gamma=\alpha\exp\left[-\frac{\omega}{T_{H}}\left(1-\frac{\omega}{2M}\right)\right],\label{eq: Parikh Correction}
\end{equation}
where $\alpha\sim1$ and the additional term $\frac{\omega}{2M}\:$
is present. By introducing the \emph{effective temperature }\cite{key-13,key-14}
\begin{equation}
T_{E}(\omega)\equiv\frac{2M}{2M-\omega}T_{H}=\frac{1}{4\pi\left(2M-\omega\right)},\label{eq: Corda Temperature}
\end{equation}
Eq. (\ref{eq: Parikh Correction}) can be rewritten in a Boltzmann-like
form \cite{key-13,key-14}, namely
\begin{equation}
\Gamma=\alpha\exp[-\beta_{E}(\omega)\omega]=\alpha\exp\left(-\frac{\omega}{T_{E}(\omega)}\right),\label{eq: Corda Probability}
\end{equation}
where $\exp[-\beta_{E}(\omega)\omega]$ is the \emph{effective Boltzmann
factor,} with \cite{key-13,key-14} 
\begin{equation}
\beta_{E}(\omega)\equiv\frac{1}{T_{E}(\omega)}.\label{eq: beta E}
\end{equation}
Therefore, the effective temperature replaces the Hawking temperature
in the equation of the probability of emission as dynamical quantity.
There are indeed various fields of science where one can take into
account the deviation from the thermal spectrum of an emitting body
by introducing an effective temperature which represents the temperature
of a black body that would emit the same total amount of radiation\emph{
}\cite{key-13,key-14}\emph{.} The effective temperature depends on
the energy-frequency of the emitted radiation and the ratio $\frac{T_{E}(\omega)}{T_{H}}=\frac{2M}{2M-\omega}$
represents the deviation of the BH radiation spectrum from the strictly
thermal feature due to the BH dynamical behavior \cite{key-13,key-14}.
Besides, one can introduce other
\emph{effective quantities}. In particular, if $M$ is the initial
BH mass \emph{before} the emission, and $M-\omega$ is the final BH
mass \emph{after} the emission, the \emph{BH} \emph{effective mass
}and the \emph{BH effective horizon }can be\emph{ }introduced as \cite{key-13,key-14}
\begin{equation}
M_{E}\equiv M-\frac{\omega}{2},\mbox{ }r_{E}\equiv2M_{E}.\label{eq: effective quantities}
\end{equation}
They represent the BH mass and horizon\emph{ during} the BH
contraction, i.e. \emph{during} the emission of the particle \cite{key-13,key-14}, respectively.
These are average quantities. The variable \emph{$r_{E}$
}is indeed the average of the initial and final horizons while \emph{$M_{E}$
}is the average of the initial and final masses \cite{key-13,key-14}.
In regard to the effective temperature, it is the inverse of the average value of the inverses of the initial and final Hawking temperatures; \emph{before}
the emission we have $T_{H}^{i}=\frac{1}{8\pi M}$, \emph{after}
the emission $T_{H}^{f}=\frac{1}{8\pi(M-\omega)}$ \cite{key-13,key-14}.
To show that the effective mass is indeed the correct
quantity which characterizes the BH dynamical behavior, one can rely on
Hawking's periodicity argument \cite{key-16,key-17,key-18}. One rewrites Eq. (\ref{eq: beta E})
as \cite{key-20} 
\begin{equation}
\beta_{E}(\omega)\equiv\frac{1}{T_{E}(\omega)}=\beta_{H}\left(1-\frac{\omega}{2M}\right),\label{eq: beta E-1}
\end{equation}
where $\beta_{H}\equiv\frac{1}{T_{H}}$. Following Hawking' s arguments
\cite{key-16,key-17,key-18}, the Euclidean form of the metric is given by \cite{key-20} 
\begin{equation}
ds_{E}^{2}=x^{2}\left[\frac{d\tau}{4M\left(1-\frac{\omega}{2M}\right)}\right]^{2}+\left(\frac{r}{r_{E}}\right)^{2}dx^{2}+r^{2}(\sin^{2}\theta d\varphi^{2}+d\theta^{2}).\label{eq: euclidean form}
\end{equation}
This equation is regular at $x=0$ and $r=r_{E}$. One also treats
$\tau$ as an angular variable with period $\beta_{E}(\omega)$ \cite{key-13,key-14,key-16}. Following \cite{key-20}, one replaces the quantity $\sum_{i}\beta_{i}\frac{\hslash^{i}}{M^{2i}}$
in \cite{key-18} with $-\frac{\omega}{2M}.$ Then, following the analysis presented in \cite{key-18}, one
obtains \cite{key-20} 
\begin{equation}
ds_{E}^{2}\equiv\left(1-\frac{2M_{E}}{r}\right)dt^{2}-\frac{dr^{2}}{1-\frac{2M_{E}}{r}}-r^{2}\left(\sin^{2}\theta d\varphi^{2}+d\theta^{2}\right).\label{eq: Hilbert effective}
\end{equation}
One can also show that $r_{E}$ in Eq. (\ref{eq: euclidean form})
is the same as in Eq. (\ref{eq: effective quantities}). 

Despite the above analysis has been realized for emissions of particles,
one immediately argues by symmetry that the same analysis works also
in the case of absorptions of external particles, which can be considered
as emissions having opposite sign. Thus, the effective quantities
(\ref{eq: effective quantities}) become 
\begin{equation}
M_{E}\equiv M+\frac{\omega}{2},\mbox{ }r_{E}\equiv2M_{E}.\label{eq: effective quantities absorption}
\end{equation}
Now they represents the BH mass and horizon\emph{ during}
the BH expansion, i.e., \emph{during} the absorption of the particle, respectively.
Hence, Eq. (\ref{eq: Hilbert effective}) implies that, in order to
take the BH dynamical behavior into due account, one must replace
the BH mass $M$ with the BH effective mass $M_{E}$ in Eqs. (\ref{eq: energia potenziale BH}),
(\ref{eq: Schrodinger equation BH}), (\ref{eq: spettro energia BH}), (\ref{eq: energia totale 2}), obtaining 
\begin{equation}
V(r)=-\frac{M_{E}^{2}}{r},\label{eq: energia potenziale BH effettiva}
\end{equation}
\begin{equation}
-\frac{1}{2M_{E}}\left(\frac{\partial^{2}X}{\partial r^{2}}+\frac{2}{r}\frac{\partial X}{\partial r}\right)+VX=EX,\label{eq: Schrodinger equation BH effettiva}
\end{equation}
\begin{equation}
E_{n}=-\frac{r_{E}^{2}M_{E}^{3}}{8n^{2}},\label{eq: spettro energia BH effettivo}
\end{equation}
\begin{equation}
E=-\frac{M_{E}}{2}.\label{eq: energia totale effettiva}
\end{equation}
From the quantum point of view, we want to obtain the energy
eigenvalues as being absorptions starting from the BH formation, that
is from the BH having null mass. This implies that we must replace
$M\rightarrow0$ and $\omega\rightarrow M$ in Eq. (\ref{eq: effective quantities absorption}).
Thus, we obtain 
\begin{equation}
M_{E}\equiv\frac{M}{2},\mbox{ }r_{E}\equiv2M_{E}=M.\label{eq: effective quantities absorption finali}
\end{equation}
Following again \cite{key-5}, one inserts Eqs. (\ref{eq: energia totale effettiva})
and (\ref{eq: effective quantities absorption finali}) into Eq. (\ref{eq: spettro energia BH effettivo}),
obtaining the BH mass spectrum as 
\begin{equation}
M_{n}=2\sqrt{n},\label{eq: spettro massa BH finale}
\end{equation}
and by using Eq. (\ref{eq: energia totale effettiva}) one finds the
BH energy levels as 
\begin{equation}
E_{n}=-\frac{1}{2}\sqrt{n}.\label{eq: BH energy levels finale.}
\end{equation}
Remarkably, in its absolute value, this final result is consistent with the
BH energy spectrum which was conjectured by Bekenstein in 1974 \cite{key-7}. Bekenstein indeed obtained $E_{n}\sim \sqrt{n}$ by using the
Bohr-Sommerfeld quantization condition because he argued that the
Schwarzschild BH behaves as an adiabatic invariant. Besides, Maggiore \cite{key-8}
conjectured a quantum description of BH in terms of quantum membranes.
He obtained the energy spectrum 
\begin{equation}
E_{n}=\sqrt{\frac{A_{0}n}{16\pi}}.\label{eq: spettro massa buco nero membrane}
\end{equation}
One sees that, in its absolute value, the
result of Eq. (\ref{eq: BH energy levels finale.}) is consistent
also with Maggiore's result. On the other hand, it should be noted that both Bekenstein and
Maggiore used heuristic analyses, approximations and/or conjectures.
Instead, Eq. (\ref{eq: spettro massa buco nero}) has been obtained
through an exact quantization process. In addition, neither Bekenstein
nor Maggiore realized that the BH energy spectrum must have negative
eigenvalues because the ``gravitational hydrogen atom'' is a quantum
system composed by bound states. 

Let us again consider the analogy between the potential energy of
a hydrogen atom, given by Eq. (\ref{eq: energia potenziale atomo idrogeno}),
and the effective potential energy of the ``gravitational hydrogen
atom'' given by Eq. (\ref{eq: energia potenziale BH effettiva}).
Eq. (\ref{eq: energia potenziale atomo idrogeno}) represents
the interaction between the nucleus of the hydrogen atom, having a
charge $e$ and the electron, having a charge $-e.$ Eq. (\ref{eq: energia potenziale BH effettiva})
represents the interaction between the nucleus of the ``gravitational
hydrogen atom'', i.e. the BH, having an effective, dynamical mass
$M_{E}$, and another, mysterious, particle, i.e., the ``electron''
of the ``gravitational hydrogen atom'' having again an effective,
dynamical mass $M_{E}$. Therefore, let us ask: what is the ``electron''
of the BH? An intriguing answer to this question has been given by
one of the authors (CC), who recently developed a semi-classical Bohr-like
approach to BH quantum physics where, for large values of the principal
quantum number $n,$ the BH quasi-normal modes (QNMs), \textquotedblleft triggered\textquotedblright{}
by emissions (Hawking radiation) and absorption of external particles,
represent the \textquotedblleft electron\textquotedblright{} which
jumps from a level to another one; the absolute values of the
QNMs frequencies represent the energy \textquotedblleft shells\textquotedblright{}
of the \textquotedblleft gravitational hydrogen atom\textquotedblright.
In this context, the QNM jumping from a level to another one has been
indeed interpreted in terms of a particle quantized on a circle \cite{key-13,key-14},
which is analogous to the electron travelling in circular orbits around
the hydrogen nucleus, similar in structure to the solar system, of
Bohr's semi-classical model of the hydrogen atom \cite{key-21,key-22}.
Therefore, the results in the present paper seem consistent with the above mentioned works \cite{key-13,key-14}.

For the BH ground state ($n=1$), from Eq. (\ref{eq: spettro massa BH finale})
one gets the mass as 
\begin{equation}
M_{1}=2,\label{eq: massa minima}
\end{equation}
in Planck units. Thus, in standard units one gets $M_{1}=2m_{P},$
where $m_{P}$ is the Planck mass, $m_{P}=2,17645\text{\texttimes}10^{-8}\mathrm{~Kg}.$
To this mass is associated a total negative energy arising from Eq.
(\ref{eq: BH energy levels finale.}), which is 
\begin{equation}
E_{1}=-\frac{1}{2},\label{eq: energia minima}
\end{equation}
and a Schwarzschild radius
\begin{equation}
r_{g1} = 4.    
\end{equation}
Hence, this is the state having minimum mass and minimum energy (the
energy of this state is minimum in absolute value; in its real value,
being negative, it is maximum). In other words, this ground state
represents the smallest possible BH. In the case of Bohr's semi-classical
model of hydrogen atom, the Bohr radius, which represents the classical
radius of the electron at the ground state, is \cite{key-6} 
\begin{equation}
\text{Bohr radius}=b_1=\frac{1}{m_{e}e^{2}},\label{eq: Bohr radius}
\end{equation}
where $m_{e}$ is the electron mass. To obtain the correspondent
``Bohr radius'' for the ``gravitational hydrogen atom'', one needs
to replace both the electron mass $m_{e}$ and the charge $e$ in Eq. (\ref{eq: Bohr radius})
with the effective mass of the BH ground state, which is $\frac{M_{1}}{2}=1.$
Thus, now the ``Bohr radius'' becomes
\begin{equation}
b_{1}=1,\label{eq: Bohr radius-1}
\end{equation}
which in standard units reads $b_{1}=l_{P},$ where $l_{P}=1,61625\text{\texttimes}10^{-35}\mathrm{~m}$
is the Planck length. Hence, we have found that the ``Bohr radius''
associated to the smallest possible black hole is equal to the Planck length. Following \cite{key-5}, the wave-function associated to the BH ground state is
\begin{equation}
\Psi_{1}=2b_{1}^{-\frac{3}{2}}r\exp\left(-\frac{r}{b_{1}}\right)=2r\exp\left(-r\right),\label{eq: wave-function 1 BH}
\end{equation}
where $\Psi_{1}$ is normalized as
\begin{equation}
\int_{0}^{\infty}\Psi_{1}^{2}dr=1.\label{eq: normalizzazione BH}
\end{equation}
The size of this BH is of the order of
\begin{equation}
\bar{r}_{1}=\int_{0}^{\infty}\Psi_{1}^{2}rdr=\frac{3}{2}b_{1}=\frac{3}{2}.\label{eq: size BH 1}
\end{equation}
The issue that the size of the BH ground state is, on average, shorter
than the gravitational radius could appear surprising, but one recalls
again that one interprets the ``BH electron states'' in terms of
BH QNMs \cite{key-13,key-14}. Thus, the BH size which is, on average,
shorter than the gravitational radius, seems consistent with the issue
that the BH horizon oscillates with damped oscillations when the BH
energy state jumps from a quantum level to another one through emissions
of Hawking quanta and/or absorption of external particles.

This seems an interesting quantum representation of the Schwarzschild
BH ground state at the Planck scale. This Schwarzschild BH ground
state represents the BH minimum energy level which is compatible with
the generalized uncertainty principle (GUP) \cite{key-12}. The GUP
indeed prevents a BH from its total evaporation by stopping Hawking's
evaporation process in exactly the same way that the usual uncertainty
principle prevents the hydrogen atom from total collapse \cite{key-12}. 

Now, let us discuss a fundamental issue. Can one say that the quantum
BH expressed by the system of equations from (\ref{eq: energia potenziale BH effettiva}) to (\ref{eq: energia totale effettiva}) is non-singular? It seems
that the correct answer is yes. It is well known that, in the classical
general relativistic framework, in the internal geometry all time-like
radial geodesics of the collapsing star terminate after a lapse of
finite proper time in the termination point $r=0$ and it is impossible
to extend the internal space-time manifold beyond that termination
point \cite{key-1}. Thus, the point $r=0$ represents a singularity
based on the rigorous definition by Schmidt \cite{key-37}. But what
happens in the quantum framework that has been analysed in this paper
is completely different. By inserting the constraints for a Schwarzschild
BH in Rosen's quantization process applied to the gravitational collapse,
it has been shown that the completely collapsed object has been split
in a two-particle system where the two components strongly interact
with each other through a quantum gravitational interaction. In concrete
terms, the system that has been analysed is indeed formally equal
to the well known system of two quantum particles having finite distance
with the mutual attraction of the form $1/r$ \cite{key-6}. These
two particles are the ``nucleus'' and the ``electron'' of the
``gravitational hydrogen atom''. Thus, the key point is the meaning
of the word ``particle'' in a quantum framework. Quantum particles
remain in an uncertain, non-deterministic, smeared, probabilistic
wave-particle orbital state \cite{key-6}. Then, they cannot be localized
in a particular ``termination point where all time-like radial geodesics
terminate''. As it is well known, such a localization is also in
contrast with the Heisenberg uncertainty principle (HUP). The HUP says indeed that either the location
or the momentum of a quantum particle such as the BH ``electron''
can be known as precisely as desired, but as one of these quantities
is specified more precisely, the value of the other becomes increasingly
indeterminate. This is not simply a matter of observational difficulty,
but rather a fundamental property of nature. This means that,
within the tiny confines of the ``gravitational atom'', the ``electron''
cannot really be regarded as a ``point-like particle''
having a definite energy and location. Thus, it is somewhat misleading
to talk about the BH ``electron'' ``falling into''
the BH ``nucleus''. In other words, the Schwarzschild radial coordinate
cannot become equal to
zero. The GUP makes even stronger this last statement: as we can notice from its general expression \cite{key-38} 
\begin{equation}
\Delta x\Delta p\geq\frac{1}{2}\left[1+\eta\left(\Delta p\right)^{2}+\ldots \right],\label{eq: GUP}
\end{equation}
it implies a non-zero lower bound on the minimum value of the uncertainty on the particle's position $\left(\Delta x\right)$ which is of order of the Planck
length \cite{key-38}. In other words, the GUP implies the existence
of a minimal length in quantum gravity. 

One notes also another important difference between the hydrogen
atom of quantum mechanics \cite{key-6} and the ``gravitational hydrogen
atom'' discussed in this paper. In the standard hydrogen atom the
nucleus and the electron are different particles. In the quantum BH analysed here they are equal particles instead, as one easily checks
in the system of equations from (\ref{eq: energia potenziale BH effettiva})
to (\ref{eq: energia totale effettiva}). Thus, the ``nucleus''
and the ``electron'' can be mutually exchanged without varying the
physical properties of the system. Hence, the quantum state of the
two particles seems even more uncertain, more non-deterministic, more
smeared and more probabilistic than the corresponding quantum states
of the particles of the hydrogen atom. These quantum argumentations
seem to be strong argumentations in favour of the non-singular behavior
of the Schwarzschild BH in a quantum framework. Notice that the results
in this paper are also in agreement with the general conviction that
quantum gravity effects become fundamental in the presence of strong
gravitational fields. In a certain sense, the results in this paper
permit to ``see into'' the Schwartzschild BH. The authors hope to
further deepen these fundamental issues in future works.

\section{Area quantization}

Bekenstein proposed that the area
of the BH horizon is quantized in units of the Planck length in quantum
gravity (let us remember that the Planck length
is equal to one in Planck units) \cite{key-7}. His result was that the Schwarzschild
BH area quantum is $\Delta A=8\pi$ \cite{key-7}.

In the Schwarzschild
BH the \emph{horizon area} $A$ is related to the
mass by the relation $A=16\pi M^{2}.$ Thus, a variation $\Delta M\,$ of the mass implies a variation
\begin{equation}
\Delta A=32\pi M\Delta M\label{eq: variazione area}
\end{equation}
of the area. Let us consider a BH which is excited at the level $n$. The corresponding BH mass is given by Eq. (\ref{eq: spettro massa BH finale}),
that is 
\begin{equation}
M_{n}=2\sqrt{n}.\label{eq: spettro massa BH finale n}
\end{equation}
Now, let us assume that a neighboring particle is captured by the
BH causing a transition from $n$ to $n+1.$ Then, the variation of the BH mass is 
\begin{equation}
M_{n+1}-M_{n}=\Delta M_{n\rightarrow n+1},\label{eq: absorbed}
\end{equation}
where 
\begin{equation}
M_{n+1}=2\sqrt{n+1}.\label{eq: spettro massa BH finale n+1}
\end{equation}
Therefore, using Eqs. (\ref{eq: variazione area}) and (\ref{eq: absorbed})
one gets
\begin{equation}
\Delta A_{n}\equiv32\pi M_{n}\Delta M_{n\rightarrow n+1}.\label{eq: area quantum a}
\end{equation}
Eq. (\ref{eq: area quantum a}) should give the area quantum
of an excited BH when one considers an absorption from the level $n$
to the level $n+1$ in function of the principal quantum number $n$.
But, let us consider the following problem. An emission from the level
$n+1$ to the level $n$ is now possible due to the potential emission
of a Hawking quantum. Then, the correspondent mass lost by the BH
will be 
\begin{equation}
M_{n+1}-M_{n}=-\Delta M_{n\rightarrow n+1}\equiv\Delta M_{n+1\rightarrow n}.\label{eq: emitted}
\end{equation}
Hence, the area quantum for the transition (\ref{eq: emitted}) should
be 
\begin{equation}
\Delta A_{n}\equiv32\pi M_{n+1}\Delta M_{n+1\rightarrow n},\label{eq: area quantum e}
\end{equation}
and one gets the strange result that the absolute value of the area
quantum for an emission from the level $n+1$ to the level $n\;$
is different from the absolute value of the area quantum for an absorption
from the level $n$ to the level $n+1$ because it is $M_{n+1}\neq M_{n}$.
One expects the area spectrum to be the same for absorption and emission
instead. In order to resolve this inconsistency, one considers the \emph{effective
mass,} which has
been introduced in Section 3, corresponding to the transitions between the two levels
$n$ and $n+1$. In fact, the effective mass is the same for emission
and absorption 
\begin{equation}
M_{E(n,\;n+1)}\equiv\frac{1}{2}\left(M_{n}+M_{n+1}\right) = \sqrt{n}+\sqrt{n+1}.
\label{eq: massa effettiva n}
\end{equation}
By replacing $M_{n+1}$ with $M_{E(n,\;n+1)}$ in equation (\ref{eq: area quantum e})
and $M_{n}$ with $M_{E(n,\;n+1)}$ in Eq. (\ref{eq: area quantum a})
one obtains 
\begin{equation}
\begin{array}{c}
\Delta A_{n+1}\equiv32\pi M_{E(n,\;n+1)}\,\Delta M_{n+1\rightarrow n}\qquad \text{emission}\\
\\
\hspace{0.75cm}\Delta A_{n}\equiv32\pi M_{E(n,\;n+1)}\,\Delta M_{n\rightarrow n+1}\qquad \text{absorption}
\end{array}\label{eq: expects}
\end{equation}
and now it is $|\Delta A_{n}|=|\Delta A_{n-1}|.$ By using Eqs.
(\ref{eq: absorbed}) and (\ref{eq: massa effettiva n}) one finds 
\begin{equation}
|\Delta A_{n}|=|\Delta A_{n+1}|=64\pi,\label{eq: 8 pi planck}
\end{equation}
which is similar to the original result found by Bekenstein in 1974 \cite{key-7},
but with a different coefficient. This is not surprising because
there is no general consensus on the area quantum. In fact, in \cite{key-27,key-28}
Hod considered the black hole QNMs like quantum levels
for absorption of particles, obtaining a different numerical
coefficient. On the other hand, the expression found by Hod
\begin{equation}
\Delta A=4\ln3
\end{equation}
is actually a special case of the one suggested by Mukhanov in \cite{key-29}, who proposed
\begin{equation}
\Delta A=4\ln k, \quad k=2,3,\ldots    
\end{equation}
This can be found in \cite{key-9,key-30}.

Thus, the approach in this paper seems consistent with the
Bekenstein area law.

\section{Discussion and conclusion remarks}

Rosen's quantization approach has been applied to the gravitational
collapse in the simple case of a pressureless \textquotedblleft \emph{star
of dust}\textquotedblright. In this way, the gravitational potential,
the Schr\"odinger equation and the solution for the collapse's energy
levels have been found. After that, by applying the constraints for
a BH and by using the concept of BH\emph{ }effective state \cite{key-13,key-14},
the analogous results and the energy spectrum have been
found for the Schwarzschild BH. Remarkably, such an energy spectrum
is consistent (in its absolute value) with both the one which was found
by J. Bekenstein in 1974 \cite{key-7} and that found
by Maggiore in \cite{key-8}.
The discussed approach also allowed to find an interesting quantum
representation of the Schwarzschild BH ground state at the Planck
scale; in other words, the smallest BH has been found, by also showing
that it has a mass of two Planck masses and a ``Bohr radius'' equal to the 
Planck length. Furthermore, two fundamental issues such as Bekenstein area law and singularity resolution have been discussed. Thus, despite the gravitational collapse analysed in this paper is the simplest possible, the analysis that it has been performed permitted to obtain important results in BH quantum physics.

Finally, for the sake of completeness, it is necessary to discuss
how the results found in this paper are related to those appearing in the
literature. The results of Bekenstein and Maggiore
concerning the BH energy spectrum have been previously cited. In general, such an energy spectrum has been discussed and derived
in many different ways, see for example \cite{key-31,key-32}. In the so-called reduced phase space quantization method
\cite{key-31,key-32}, the BH energy spectrum gets augmented by an
additional zero-point energy; this becomes important if one attempts
to address the ultimate fate of BH evaporation, but, otherwise, it
can safely be ignored for macroscopic BHs for which the principal
quantum number $n$ will be extremely large. It is important to stress
that the discreteness of the energy spectrum needs a drastic departure
from the thermal behavior of the Hawking radiation spectrum. A popular way to realize this is through
the popular tunnelling framework arising from the paper of Parikh
and Wilczek \cite{key-17}. In that case, the energy conservation
forces the BH to contract during the emission of the particle; the horizon recedes from its original radius and becomes smaller at the end of the emission process \cite{key-17}.
Therefore, BHs do not exactly emit like perfect black bodies,
see also the discussion on this issue in Section 3. Moreover, Loop Quantum
Gravity predicts a discrete energy spectrum which indicates a physical
Planck scale cutoff of the Hawking temperature law \cite{key-33}.
In the framework of String-Theory one can identify microscopic BHs with long chains
living on the worldvolume of two dual Euclidean brane pairs \cite{key-34}.
This leads to a discrete Bekenstein-like energy spectrum for
the Schwarzschild black hole \cite{key-35}. The Bekenstein energy spectrum
is present also in canonical quantization schemes \cite{key-36}. This approach
yields a BH picture that is shown to be equivalent to a collection
of oscillators whose density of levels is corresponding to that of the statistical bootstrap model \cite{key-36}.

In a series of interesting papers \cite{key-37,key-38,key-39}, Stojkovic
and collaborators wrote down the Schr\"odinger equation for a collapsing
object and showed by explicit calculations that quantum mechanics
is perhaps able to remove the singularity at the BH center (in various 
space-time slicings). This is consistent with our analysis. In \cite{key-37,key-38,key-39} it is indeed proved (among other things) 
that the wave function of the collapsing object is non-singular at the center 
even when the radius of the collapsing object (classically) reaches zero. 
Moreover, in \cite{key-39}, they also considered charged BHs.

Another interesting approach to the area quantization, based on graph
theory, has been proposed by Davidson in [41]. In such a paper, the Bekenstein-Hawking
area entropy formula is obtained, being automatically accompanied
by a proper logarithmic term (a subleading correction), and the size
of the horizon unit area is fixed \cite{key-40}. Curiously, Davidson
also found a hydrogen-like spectrum in a totally different contest \cite{key-41}.

Thus, it seems that the results of this paper are consistent
with the previous literature on BH quantum physics.

\section*{Acknowledgements }
The authors thank Dennis Durairaj for useful suggestions. The authors are also grateful to the anonymous referee for very useful comments.

\providecommand{\href}[2]{#2}

\address{
International Institute for Applicable Mathematics\\
and Information Sciences (IIAMIS), B.M. Birla Science\\ 
Centre Adarsh Nagar, Hyderabad - 500 463, India\\ 
and Dipartimento di Matematica e Fisica, Istituto Livi\\
Via Antonio Marini, 9,59100 Prato (Italy)\\ 
\email{cordac.galilei@gmail.com}\\
\\
Institute for Theoretical Physics, Utrecht University\\
Princetonplein 5, 3584 CC Utrecht, The Netherlands\\
\email{feleppa.fabiano@gmail.com}\\
}

\end{document}